\documentclass[iop]{emulateapj}
\usepackage[usenames, dvipsnames]{color}
\usepackage{amsmath}
\usepackage{graphicx}
\usepackage{rotating} 

\shorttitle{VVV Survey Microlensing}
\shortauthors{Navarro et al. 2020}

\begin{document}
\title{VVV Survey Microlensing: Candidate Events with Source in the Far Disk}

\author{Mar\'ia Gabriela Navarro \altaffilmark{1,2,3,*} 
Dante Minniti \altaffilmark{1,3,4} 
Rodrigo Contreras Ramos \altaffilmark{3,5}}

\affil{$^1$Departamento de Ciencias F\'isicas, Facultad de Ciencias Exactas, Universidad Andr\'es Bello, Av. Fernandez Concha 700, Las Condes, Santiago, Chile}
\affil{$^2$Dipartimento di Fisica, Universita di Roma La Sapienza, P.le Aldo Moro, 2, I-00185 Rome, Italy}
\affil{$^3$Millennium Institute of Astrophysics, Av. Vicuna Mackenna 4860, 782-0436, Santiago, Chile}
\affil{$^4$Vatican Observatory, V00120 Vatican City State, Italy}
\affil{$^5$Instituto de Astrofisica, Pontificia Universidad Cat\'olica de Chile, Av. Vicuna Mackenna 4860, 782-0436 Macul, Santiago, Chile}

\begin{abstract}
The VVV microlensing search has recently surveyed the region of the Galactic plane at $b=0$ within $-10.00 \leq l \leq 10.44$ deg. in the near-infrared (IR), discovering hundreds of microlensing events.
In this paper we explore the microlensing events with background sources that might be located in the far disk of the Galaxy, beyond the bulge. 
We discuss the possible configurations for the microlensing lenses and sources located at different places within the Galactic plane.
Then we search for these events using the local red clump centroids of the VVV near-IR color-magnitude diagrams.
According to the estimated distances and proper motions, $N=20$ events may have sources located in the far disk.
The candidates for far-disk sources show on average longer timescales ($t_E= 49.3 \pm 7.9$ days) than the mean of the timescale distribution for the bulge red clump sources ($t_E= 36.4 \pm 1.1$ days). 
We conclude that the population of microlensing events in the region $-10.00 \leq l \leq 10.44$, $-0.46  \leq b \leq 0.65 $ deg. contains a non-negligible number of events with candidate far-disk sources ($\sim 11 \%$).
Our results are relevant in view of the future microlensing plans with the Roman Space Telescope (formerly WFIRST) in the near-IR. 
\end{abstract}


\section{Introduction}
\footnotetext{Corresponding author: mariagabriela.navarro$@$uniroma1.it}

In the framework of the study of gravitational lenses within the Galaxy, Milky Way (MW) microlensing models and observations are incomplete, because previously they did not take into account or detect the other half of the Milky Way, beyond the Galactic bulge. Distant sources for the microlensing events were considered only for the case of sources in the Sgr dwarf galaxy (e.g. \cite{Alcock97,Cse2001}), but sources located in the far disk of the MW, lying beyond the bulge (with distances $D \gtrsim 10$ kpc), have been largely ignored. 
 These events in the far disk may be numerous in the MW plane, but current optical microlensing searches are relatively insensitive to them because they explore higher Galactic latitudes. 
  
 In fact, the existing models  (e.g. \cite{Han95,Han03,wood05,ryu08,penny13, penny19,Henderson14,awiphan16,Henderson16, Poleski16,wegg16}) do not consider the events with sources in the far disk of the Milky Way.
This is understandable because previous microlensing surveys were carried out at optical wavelengths, exploring regions relatively far from the Galactic plane (where dust obscuration blinds their search). However, the maximum number of microlensing events is expected to occur in the Galactic plane, including in particular the microlensing events with sources in the far disk, which would be located at very low latitudes, where crowding and extinction are also maximal.

The detection of microlensing events with sources in the far disk and the subsequent analysis of the frequency of these events with respect to events with sources in the Galactic bulge is especially interesting for studies of Galactic structure and dynamics.

Near-infrared (IR) microlensing surveys have the potential to observe through the dust in the Galactic plane at $|b| \sim 0$ and reach far beyond the Galactic bulge. 
An encouraging development in this vein was the first discovery of a microlensing event with a source on the other side of the Galaxy beyond the Galactic bulge, which was recently made by the UKIRT microlensing survey \citep{y18}.  
Likewise, \cite{bennett18b} reported the discovery of a microlensing event at low latitude. One of the possible configurations discussed in this article is that the star is a subgiant star located on the other side of the Galactic bulge.

Ground-based near-IR microlensing surveys like the UKIRT \citep{Shvartzvald17}, and the VVV \citep{navarro17,navarro18, navarro20b} are well-suited for piercing through the Milky Way disk, unveiling its unexplored most distant regions. As argued by \cite{Shvartzvald17} and \cite{navarro17,navarro18, navarro20b}, these near-IR ground-based surveys are pathfinders for the future Nancy Grace Roman Space Telescope (hereafter Roman), formerly known as the Wide-Field InfraRed Survey Telescope or the WFIRST microlensing survey (\cite{Spergel15,bennett18,penny19}). 

The VVV survey in particular has been mapping the inner bulge and southern disk since 2010, and the near-IR photometry and the image resolution are good enough to see structures well beyond the bulge even at low latitudes \citep{gonzalez18}, and also to search for microlensing events with sources in the far disk. 

The VVV survey microlensing project started searching the Galactic center region (tiles $b332$, $b333$, and $b334$), finding $182$ events \citep{navarro17}. We then expanded the area searching for events across the plane at Galactic latitude $b=0$ deg., finding $630$ events (of which $291$ are red clump events) between Galactic longitudes $-10.00<l<10.44$ deg. \citep{navarro18, navarro20b}. 
We found asymmetry in the distribution across Galactic longitudes, in the sense that there are about $25\%$ more events at positive longitudes. 
This asymmetric effect was predicted by the existing models (e.g. \cite{Han03}), as a consequence of the barred bulge.
In \cite{navarro20a} we explored the latitude dependence of microlensing, finding $N = 238$ new events along the Galactic minor axis with latitudes $-3.7<b<3.9$ deg.
We found a steep increase in the number of events toward the Galactic plane at $b=0$ deg., much steeper than any expectations from the previous optical surveys and models.
We suggested that this may be due to the presence of far-disk microlensing events.

In this paper we explore the distance distribution of the microlensing sources in more detail, using red clump (RC) giants as tracers.
First, we define the different kinds of events with sources in the far disk that have different spatial distributions and kinematics (Section  \ref{sec:sec2}). 
Then we identify specific candidates from the RC sample of \cite{navarro18, navarro20b} that could correspond to far-disk sources (Section \ref{sec:sec3}).
In Section  \ref{sec:sec4} we discuss the different models and detections related to this population.
We discuss the results, and we suggest a new strategy for the future Roman Space Telescope observations (Section  \ref{sec:sec5}).
We present our conclusions are presented in Section \ref{sec:sec6}.

\section{Definition of events}
\label{sec:sec2}
In the past it was standard to consider three types of Galactic microlensing events toward the bulge (e.g. \cite{gould95}):

$\bullet$ Bulge -- disk events: with sources in the bulge  ($D\sim 8.3 \pm 2$ kpc, \cite{dek13}) and lenses in the foreground disk ($D<6$ kpc); 

$\bullet$ Disk -- disk events: with nearby sources and lenses ($D<6$ kpc) located in the Galactic disk; and

$\bullet$ Bulge -- bulge events: with sources and lenses located in the Galactic bulge  ($D\sim 8.3 \pm 2$ kpc). \\

Past and present microlensing optical surveys did not need to be concerned about distant Galactic sources located beyond the bulge, because the fields explored were at larger $|b|$. For example, at $b=-3$ deg. in the line of sight beyond the bulge (at a distance $D>10$ kpc), the height below the plane is $z>0.5$ kpc, and the expected density of disk stars is negligible for a MW disk scale height of $h_z=0.3$ kpc. The exceptions to this are the microlensing events with sources in the Sgr dwarf galaxy, located beyond the MW disk \citep{Alcock97,Cse2001}.

In fact, previous works discussed the dependence of disk lenses and bulge lenses on the estimated source distance. This is highly ambiguous when the the detection of sources beyond the Galactic bulge becomes a reality.

The VVV survey near-IR photometry allows us to observe very low latitude fields and detect sources well beyond the bulge. In fact, \cite{gonzalez18} detected the RC of the spiral arm behind the bulge. Therefore, we consider new families of microlensing events for distant Galactic disk sources, classifying them on the basis of their different positions and kinematics \citep{navarro19}: 

$\bullet$ Far-disk -- near-disk events: where far-disk sources are those with distances beyond the Galactic bulge ($D \gtrsim 10$ kpc) and with disk kinematics, and lenses in the foreground disk ($D<6$ kpc).
These should have very high relative transverse velocities ($\Delta V_t \approx 400-500$ km/s)

$\bullet$ Far-disk -- bulge events: with sources located in the far disk ($D \gtrsim10$ kpc) and lenses located in the Galactic bulge ($D\sim 8.3 \pm 2$ kpc). These are expected to be the most numerous events with far-disk sources in our data because the stellar density along the line-of-sight peaks in the bulge. The microlensing event recently detected by \cite{y18} would be classified as a far-disk -- bulge event. 

$\bullet$ Far-disk -- far-disk events: with sources and lenses located in the far disk ($D \gtrsim 10$ kpc). Even though these sources would be faint and reddened, the number of these events would be non-negligible in our data because the effective area of the Galaxy sampled is larger. \\

Out of these three new families of microlensing events with sources in the far disk discussed here, the far-disk -- bulge events would be the most numerous. In fact, this is the most advantageous configuration for the characterization of the microlensing events, because the source distance can be estimated from the photometry/spectroscopy, and the lens distance can be assumed to be in the bulge because the density along the line of sight is always higher in the bulge (according to star counts and all Galactic models). 

\section{Identifying Candidates}
\label{sec:sec3}

\begin{figure}[t]
\epsscale{1.2}
\plotone{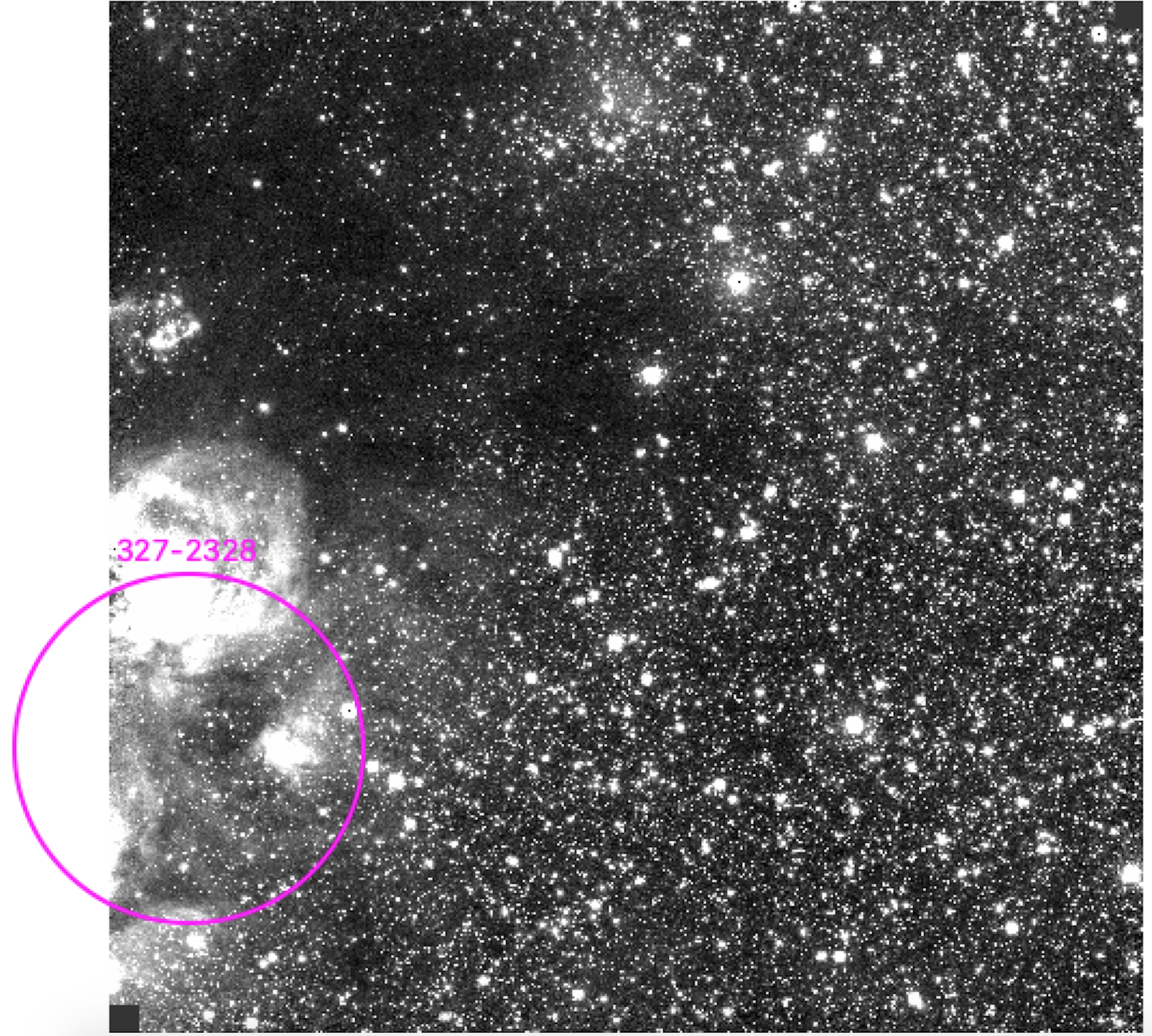}
\caption{VVV image of the event ID 2328 located in tile 327. The magenta circle has a radius of $R=2 '$. The dust and gas are irregularly distributed around the event. In this case we used the contour map obtained from \cite{gonzalez12} (shown in Figure~\ref{extinction}) to select the irregular area with the same extinction of the event itself. 
This event is not part of the final sample of candidates because it did not meet one or more of the requirements.
\label{dust}}
\end{figure}

\begin{figure}[t]
\epsscale{1.2}
\plotone{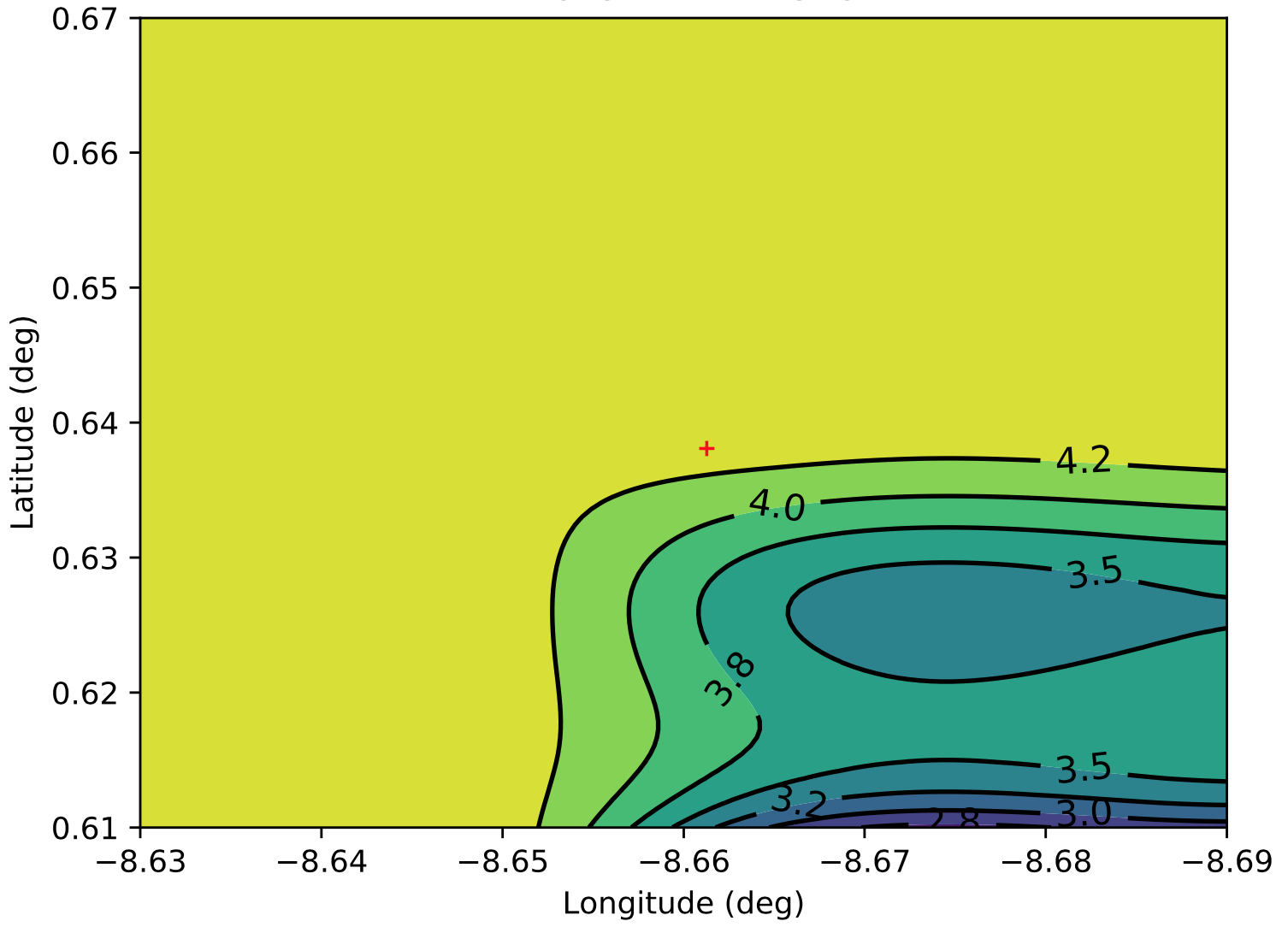}
\caption{Contour map of the region around the event  ID 2328 located in tile 327, obtained using an interpolation of the reddening maps of \cite{gonzalez12}.
In this case the values of the reddening change severely, from $E(J-K_s)=4.2$ to  $E(J-K_s)=3$ within a small area.
This event is not part of the final sample of candidates because it did not meet one or more of the requirements.
\label{extinction}}
\end{figure}

\begin{figure*}[t]
\epsscale{1}
\plotone{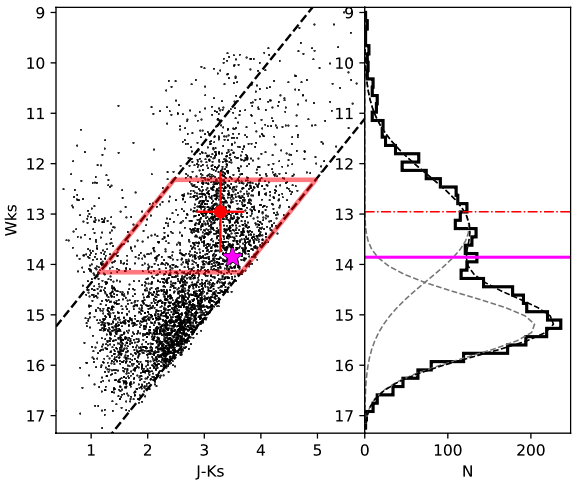}
\caption{Left panel: near-IR $W_{k_s}$ $vs.$ $J-K_s$ color-magnitude diagram for the far-disk event candidate ID 70134 located in tile 337. The black dashed lines are the limits for the area of the CMD analyzed. The red horizontal lines show the $\pm 1 \sigma$  limits of the $W_{k_s} $ distribution around the second gaussian peak.
The red circle shows the RC centroid and its dispersion for the stars located in the region within the red box. The magenta star shows the position of the far-disk microlensing source candidate. Right panel: $W_{k_s}$  histogram of the stars located within the strip limited by the black dashed lines of the left panel. The black lines show the double-Gaussian fit used to compute the RC.  The red horizontal line shows the $W_{k_s} $ of the centroid. The magenta line shows the position of the event very close to the fainter second peak that corresponds to the RC of the far disk.
\label{cmd}}
\end{figure*}

Far-disk sources would be systematically fainter and redder than bulge stars. For example, a typical RC star in the bulge has $K_s\sim 13.2$ mag and $J-K_s\sim 1.0$ mag in a low-extinction field like the Baades window with $A_V=4.5$ mag, $A_{K_s}\sim 0.5$ mag. A similar RC star would have $K_s\sim 16.0$ mag and $J-K_s\sim 6.0$ mag at the Galactic center field where the extinction is much higher ($A_V\sim 30$ mag, $A_{K_s}\sim 3$ mag). 

Detecting these sources may be possible. For example, \cite{y18} detected the first microlensing event in the far disk. They estimated a distance between $D = 10$ and $ D =14.5$ kpc depending on the method they adopt. Also, \cite{navarro18, navarro20b} argued that the observed distribution of VVV microlensing sources may be contaminated by far-disk events.

The best way to prove that we have observed far-disk sources is to estimate source distances accurately. 
This can be done for RC stars, which are good distance indicators \citep{Paczynski98} or using spectroscopy. 
Additionally, we can argue that there are a few ways to select good candidates of microlensing sources from our sample that may be located in the far disk:

1. The color-magnitude and color-color diagrams (hereafter CMDs and CCDs, respectively) can be used to probe the distant sources (especially for the RC stars), which should be fainter and more reddened than the Galactic bulge RC.

2. The proper motions (PMs) of the microlensing sources (especially the RC stars) should help to detect good candidates of sources that are located in the far disk because they should have systematically different PMs than local disk stars.

3. The microlensing event timescales with sources in the far disk might be different in general depending on the lens locations. 

Regarding the CCD and CMD analysis, \cite{gonzalez18} detected the RC from the spiral arm behind the bulge using VVV photometry. These RC stars from the spiral arm in the far disk are fainter and more reddened than the bulge giants. The magnitude difference depends on Galactic longitude, because the barred bulge is inclined with respect to the line of sight. For example, if the far-disk arm is located at a distance of about 12 kpc on average, the difference in distance modulus in magnitudes with the bulge should be
$\Delta K_s \sim 1.5$ mag at $l=-10$ deg., 
$\Delta K_s  \sim 0.8$ mag at $l=0$ deg., and 
$\Delta K_s \sim 0.0$ mag at $l=10$ deg., respectively (assuming $R_0=8.3$ kpc). 
Adopting a larger distance to the far-disk arm $D=14$ kpc, these numbers change to $\Delta K_s \sim 1.8, 1.1$, and $0.0$ mag, respectively. These are very large magnitude differences, readily measurable with the VVV photometry.
We select the most likely candidate microlensing sources from the far disk taking into account these expected magnitude differences as a function of Galactic longitude. 
In addition, if there is dust extinction between the bulge and the far arm, the distant sources would appear more reddened and even fainter. 

The main problem with detecting the far-disk candidates is the differential reddening  \citep{alonso17}. 
In order to account for this effect, the RC sources should be identified relative to their local RC centroids.
With these points in mind, we proceed to select candidates from the RC sample of \cite{navarro18} following a four-step procedure. 

The first step consisted on using the CMD to single out the stars that are fainter than the bulge RC.
We follow a procedure similar to that used in \cite{sumi13} for selecting the local RC.
We limit the area around the events with circles of radius from $R=30''$ to  $2 '$. In most of the cases the CMD remained the same but had fewer stars when we used smaller areas.
In some cases, especially in the Galactic plane, the variation of the extinction changes on scales of $<1'$.
Therefore, to deal properly with special cases of strong differential reddening or the presence of strong dust lanes, we check all the VVV images. 
An example of strong dust lanes is shown in Figure~\ref{dust}. 
In these cases we used the contour maps obtained from an interpolation of the extinction maps presented in \cite{gonzalez12} (Figure~\ref{extinction}) and computed the CMD of the stars located in areas with similar extinction values.

Then we select the area of interest where the RC stars are located. 
For this we limited the area of the CMD with a box centered in the RC region (Figure~\ref{cmd}).
The diagonal boundaries (black dashed lines of Figure~\ref{cmd}) were computed using the distribution of stars along stripes of different widths along the red color edge, then we fitted a double gaussian and selected the best strip width as the distribution with the smallest $\chi^2$ and highest difference between the peaks, representing the main sequence and RC populations. 
This procedure gave us, in some cases, the slope of the local extinction ratio.
Although we obtain different local extinction ratios for each CMD, the values varied from from  $A_{K_s}=0.3 \times E(J-K_s)$ to $A_{K_s}=0.7 \times E(J-K_s)$, for simplicity we computed the centroid using the the Wesenheit magnitude defined as 

\begin{equation}
W_{k_s} = K_s - 0.428 (J-K_s). 
\end{equation}

We obtained the $W_{k_s} $ histogram of the strip along with a double-Gaussian fit.
The limits in $W_{k_s} $ magnitudes are defined as the area of $\pm 1 \sigma$ fainter/brighter than the peak of the distribution 
($W_{k_s} $ boundaries of the box in Figure~\ref{cmd}). 
Then we measured the centroid and dispersion of the local RC using the stars located in the area within the box. 

The far-disk RC sources are expected to appear as outliers of the bulge RC. 
Therefore, if the far-disk population is detectable in the field we might identify a small overdensity of stars on the fainter and redder side of the RC. 
This overdensity should be detected as a bump in the magnitude distribution of stars.
To highlight the double bump (RC and far-disk RC) we limit the analysis to the strip computed in the previous step that excluded the near disk population, limited by the black dashed lines of Figure~\ref{cmd}.
In the cases when the far-disk RC is detectable, we expect to find it $\Delta K_s \sim 0.5-2 $ mag fainter than the main bulge RC, depending on the Galactic longitude and within the 3$\sigma$ interval in magnitude \citep{gonzalez18}.
The double bump is more evident when the CMD is corrected by extinction, therefore we used the $W_{k_s} $ magnitude distribution.
In cases of highly extincted fields, we are not able to detect far-disk stars, so the bump was not detected.

Then, to detect the far-disk source candidates we used the prior for the centroid based on two criteria.
First, we selected the CMDs with an evident presence of a far-disk population. 
If we expect to find a far-disk source in the small area we should detect other distant stars, and therefore the small bump. 
We do not consider the sources with nondetectable bumps in the CMD good far-disk candidates, although the source was located on the faint side of the RC.
In fact, the double bump, and therefore the candidates, are detected in the most central tiles that have a higher number of observations ($\sim 30 \%$) than the most external ones.
Second, we search for sources fainter and redder than the local RC, i.e. the extinction-corrected magnitude ($W_{k_s}$) that lies within the bump of the far-disk RC and is redder than the bulge RC centroid color $(J-K_s)_{cen}$ .
Due to the double peak of these cases the magnitude of the bulge RC centroid was overestimated, therefore the last step consisted of recomputing the centroid of the local RC of the bulge and its dispersion, excluding the peak of the far-disk RC stars. 
Figure~\ref{cmd} shows an example of the limits selection, centroid computed, and far-disk candidate ID 70134 located in tile 337.
A total of $N= 34$ events were selected in this step. 

We calculate the individual reddening values for the different fields on the RC centroid as $E(J-K_s) = (J-K_s)_{cen}  - (J-K_s)_0$ using the RC centroids of each CMD assuming the mean intrinsic color of the bulge RC giant stars to be $(J-K_s)_0=0.60\pm 0.01$ mag \citep{alves02}.
Note that these individual reddenings cannot be assumed to be those listed in the reddening maps of \cite{gonzalez12}, because their mean reddening values were computed for sources located at the distance of the bulge. 
As we are trying to determine which of the sources studied here are located on the far disk, we have to consider that their reddening values should be larger in principle, due to the presence of dust located beyond the bulge.

Armed with these individual reddening values, the second step is to estimate the far-disk RC star distances assuming that the extinction ratio on the far side of the disk is similar to the ratio for the dust between us and the bulge.
We adopt the mean absolute magnitude of the Galactic bulge RC as  $K_{S0}= -1.68\pm 0.03$ from \cite{alves02}.
We also use the extinction ratios measured by \cite{alonso17}: $A_{K_s}=0.428 \times E(J-K_s)$. 

We propagated the errors in every step of the calculations. 
The uncertainty in reddening $\sigma_{E(J-K_s)}= 0.31$ mag is obtained using the color error $\sigma_{(J-K_s)_0 }= 0.01 $ mag \citep{alves02} and the mean error for the centroid color $\sigma_{(J-K_s)^{cen}}= 0.31$ mag. 
Assuming that the uncertainty in extinction ratio from \cite{alonso18} is $\sim 20\%$, we obtain $\sigma_{A_{K_s}}= 0.76$ mag.
The absolute magnitude error is $\sigma_{A_{K_{s0}}}= 0.76 $ mag and using that the photometric errors are $\sigma_{K_s} = 0.01$ mag \citep{saito12, contreras17, alonso18}.
Lastly, for the distance we obtain $\sigma_d \approx 2.9$ kpc.

In this step we limit the sample to the RC stars with $D\geq 11$ kpc. 
Then the distance of each event was analyzed depending on its location, because as mentioned above, the barred bulge is angled with respect to the line of sight so the $\Delta K_s$ and mean distance to the bulge depend on the Galactic longitude. 
Therefore the distance cut for far-disk sources also depends on the longitude of the source.

As an aside, we have also estimated the reddening and distances using the new mean magnitudes and colors for the Galactic bulge RC giants from Gaia and 2MASS: $K_{S0}=-1.61 \pm 0.01$, and $(J-K_s)_0=0.66 \pm 0.01$ \citep{ruiz18}.
This produces a slight shift with respect to the mean distances estimated using the respective mean RC values from \cite{alves02}, but does not change our conclusions at all.

\begin{figure}
\epsscale{1.21}
\plotone{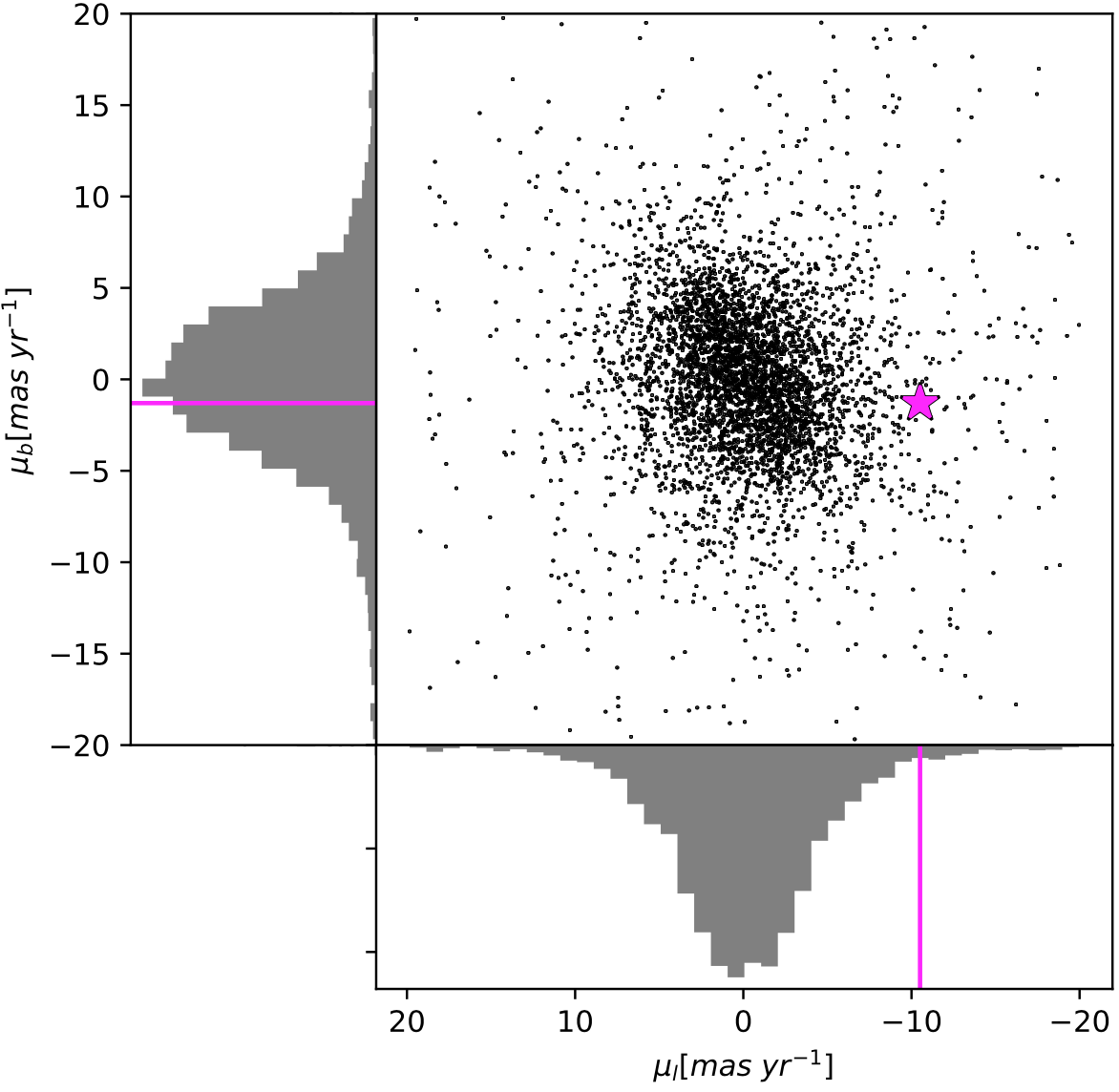}
\caption{Vector point diagram of the area around event candidate ID 90309 located in tile 334.
The magenta star shows the far-disk event candidate proper motion. 
The histograms show the distribution in each axis and the lines are the position of the far-disk event candidate. 
\label{pm}}
\end{figure}

The third step is the proper motion analysis.
The proper motions that we expect for far-disk sources candidates follow $\mu_l (mas$ $yr^{-1}) < 0$. 
Additionally, as the disk source motion is dominated by the rotation of the Galaxy, we expect a low velocity in the Galactic latitude direction, therefore we limit the sample to the events with  $| \mu_b  | (mas$ $yr^{-1}) < 4.74 \times 50 (km/s) / D (kpc)$ ($\sim |V_b| < 50 (km/s)$). 
We selected them in order to reject the very faint and red sources that can be present in the near disk and can contaminate our sample. 
To account for random motions of nearby or bulge stars that can also meet the requirements defined above, we computed the VPD for the nearby stars ($R= 2'$) and compared them with the proper motion of the source. In this step the sample was reduced to 28 events.
Figure~\ref{pm} shows one example of the local VPD around candidate ID 90309 located in tile 334.

For the catalog presented in \cite{navarro18, navarro20b}, when the source $K_s$ magnitude was not well constrained from the photometry, we used the baseline magnitude of the light curve (far from the peak). 
Additionally, when the $J$ mag was not well constrained, we used the red edge of each CMD to define the magnitude upper limit.
The fourth step for the analysis for far-disk candidates consisted of rejecting those events to avoid false positives. 
Using the same argument, we exclude from the sample the cases in which the colors and magnitudes obtained from the fitting and from the photometry were different.
Additionally, we excluded binaries and events showing parallax effect and the PSPL events that do not have well-constrained parameters or high values for the blending parameter ($f$), which affects the source flux estimation and the distance estimation. 

To be consistent with this procedure and to compute a more accurate frequency estimation we redo the RC selection of \cite{navarro18, navarro20b} considering the differential extinction, i.e., computing the RC centroids for the local CMDs around the sources. 
RC stars were selected as the sources located at 3$\sigma$ within the Gaussian fit (around 1 mag away from the main magnitude), with blending parameters consistent with zero and well constrained parameters from the standard microlensing model.
The final sample contains $N_{RC}=153$ sources.

Finally, selecting the stars that meet all the requirements mentioned above for good candidates yields $N=20$ RC sources that could be located in the far disk.
Compared with the new sample of RC stars ($N_{RC}=153$ events), the fraction of candidates for far-disk sources reaches  $\sim 11\%$ of the sample ($N_{RC}=173$ sources).

The final list of the best candidate RC stars from the far disk is presented in Table~\ref{tab:table1} along with the Galactic coordinates, magnitudes, and microlensing parameters.
Table~\ref{tab:table2} includes the proper motions, the centroid of the local field, the extinction ratio, and distances.


On the one hand, the strict and conservative selection criteria applied in this procedure to obtain the candidates could suggest that this number is a lower limit, but on the other hand there are alternative scenarios where the sources can also meet the requirements and act as false positives. 
However, the most important factors, such as anomalous extinction, blending, random motions, and the requirement of the detectable presence for the far-disk RC, were included in the analysis simultaneously.
Therefore, we can conclude that for the final sample of candidates the most plausible explanation is that the source is located in the far disk and that this type of microlensing event configuration is in fact numerous at low Galactic latitudes. 
This is as expected, since the Galactic volume explored is much larger that the near disk area sampled in front of the bulge. 


In order to estimate what timescales should be expected, we use Equation. 17 of \cite{sude12}:

\begin{equation}
\begin{aligned}
t_E \approx 38 \textrm{ day } \sqrt{\frac{D_l}{D_s}  \left( 1 - \frac{D_l}{D_s} \right) }\left( \frac{D_s}{8 kpc} \right)^{1/2} \\ \times \left( \frac{M_L}{0.3 M_\odot} \right)^{1/2} \left( \frac{V_t}{ 200 \textrm{ km s}^{-1}} \right)^{-1} ,
\end{aligned}
\end{equation}


where $M_L$ is the lens mass, $V_t$ is the relative transverse velocity, and $D_L$ and $D_S$ are the lens and source distances, respectively.


Then, for a typical far-disk source at $D_S \approx 12$ kpc and a lens in the near disk at $D_L \approx 6$ kpc, we have $V_t \approx 500$  km s$^{-1}$, and
\begin{equation}
t_E\approx 9 ~days \times \left( \frac{M}{0.3 ~M_{\odot}} \right)^{1/2}.
\end{equation}
On the other hand, for a typical far-disk source at $D_S \approx 12$ kpc and a lens in the bulge at $D_L \approx 8$ kpc, we have  $V_t \approx 250$ km s$^{-1}$, and
\begin{equation}
t_E\approx 18 ~days \times  \left( \frac{M}{0.3 ~M_{\odot}} \right)^{1/2}.
\end{equation}
However, for microlensing events with both the source and lens located in the far disk we expect longer timescales. 
For example, a far-disk source at $D_S \approx 14$ kpc and a lens in the bulge at $D_L \approx 12$ kpc, we have  $V_t \approx 25$ km s$^{-1}$, and
\begin{equation}
t_E\approx 140 ~days   \times  \left( \frac{M}{0.3 ~M_{\odot}} \right)^{1/2}.
\end{equation}

\begin{figure}
\epsscale{1.21}
\plotone{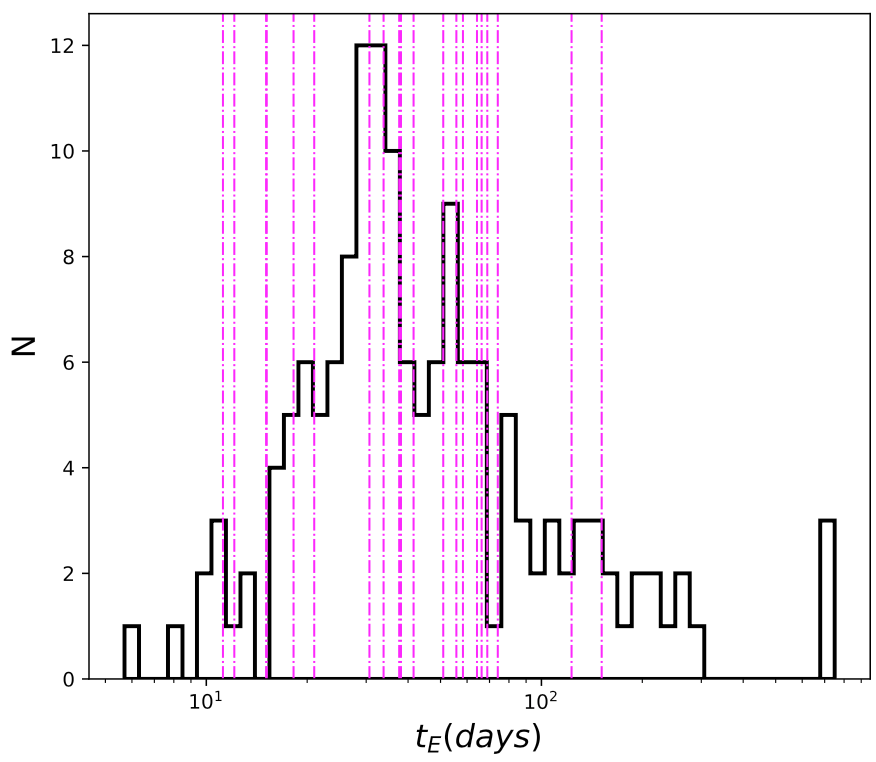}
\caption{Timescale distribution of the RC sample (black histogram). 
The magenta vertical lines show the timescale of the far-disk candidates with a mean of $t_E= 49.3 \pm 7.9$ days.
The mean timescale of the RC distribution is $t_E = 36.4 \pm 1.1$ days.
\label{t}}
\end{figure}

In this study, the final sample of far disk source candidates gives a mean timescale of $t_E= 49.3 \pm 7.9$ days. 
The standard error is large due to the small statistics, but this value is higher than the mean and the peak of the Gaussian fit obtained for the RC sample used in this analysis ($t_E= 17.3 \pm 3.8$ days and $t_E= 36.4 \pm 1.1$  respectively).
Figure~\ref{t} shows the bulge RC timescale distribution ($N_{RC}=153$ stars) and the timescale of the far-disk candidate events shown as vertical lines.
From Figure~\ref{t} is clear that, on average, far-disk candidates have longer timescales compared to the mean bulge RC distribution.

Gaussian fit for the 

\section{Existing models and detections}
\label{sec:sec4}
Microlensing optical depths for the Galactic bulge have been estimated from the optical surveys by assuming that most of the sources are located in the bulge (e.g. \cite{pop05,Hamadache06,sumi06}).
Likewise, different theoretical predictions of microlensing optical depth for the Galactic bulge have been computed by \cite{Han95,Han03,wood05,ryu08,kerins09,penny13,penny19,Henderson14,wegg16,Henderson16} and \cite{Poleski16}, among others, also assuming that the majority of the sources are located within the bulge.
Understandably, the possible presence of sources in the far disk has not been considered so far because the optical surveys avoid the high extinction regions.

The event rates in the innermost Galactic fields are unknown because these regions have not been accessible to the optical surveys, which are confined to the less reddened regions. 
The predictions from the various models for the most extincted fields in the Galactic plane show significant dispersion \citep{penny19}.
Therefore, near-IR observations are required, such as the UKIRT \citep{Shvartzvald17} and VVV surveys that can provide the first glance of this unexplored region of the Galaxy.

The microlensing models of \cite{wegg16} suggest that the inner Milky Way has a low dark matter content and a near maximal disk. The  maximal disk hypothesis says that the luminous disk alone is responsible for the circular velocity in the inner region of the rotation curve of spiral  galaxies like the Milky Way \citep{van86,Sackett97}. However, none of the existing microlensing models consider the possibility of events with sources beyond the bulge.
The detection of events with sources in the far disk should be important to constrain the maximal disk models and to study the Galactic structure and dynamics. 
Now that the near-IR surveys can reach the plane of the Galaxy, and therefore the far disk, new models are needed to contrast with our observations.
The future models need to consider the different types of events with sources in the far disk, namely far-disk -- near-disk events, far-disk -- bulge events, and far-disk --far-disk events. These models need to account for the different density distributions along the line of sight and also for the different kinematics. 


In the past the most common assumption was to place the sources at the distance of the bulge (at $D=8.3\pm 2$ kpc), and the lenses anywhere (in the foreground disk or in the bulge) along the line of sight.
As a specific example, for the microlensing events located at $l=+10$ deg., the stellar density along the line of sight peaks at the tip of the bar, at a distance of $D\sim 6$ kpc. 
This is where the lenses are going to be preferentially located if we are interested in sources situated in the far disk. 
But for $l=+10$ deg. the far disk sources are located at $D\sim 10-14$ kpc, which is nearly twice the distance of the lenses, where the lensing is most favorable.

For example, \cite{yee15}  and \cite{seb15} estimated the distances for 22 lenses in total toward the bulge, assuming that the all sources are located at $D=8.3$ kpc. The majority of them are located in the disk, with distances $1 < D < 6$ kpc, while only eight of them are located in the bulge, with $6<D<8$ kpc. In this specific case, $62\%$ are bulge-disk events, while $38\%$ are bulge-bulge events.


Particularly interesting are the microlensing events with high relative lens-source proper motions, because future high-resolution images can directly reveal the lens.
The microlensing events with sources in the far disk are more likely to exhibit high relative PMs. Among these, the far-disk--near-disk events are preferred, because their relative PMs would be the highest, but also the far-disk--bulge events would have high PMs at positive Galactic longitudes. 
There is an interesting effect expected with the longitude dependence of the relative source-lens PMs that could be studied with better resolution and a greater coverage of the area in the longitude axis.
The far-disk--bulge events would have high relative PMs at positive longitudes, while for the bulge--near-disk events the opposite is true: they would have high relative PMs at negative longitudes.
Relative proper motions can be measured only in cases where the lens is detected and therefore should be reasonably bright compared to the source.

\section{The Future with the Nancy Grace Roman Space Telescope}
\label{sec:sec5}
Even though the detection of microlensing events in the far disk counts as another achievement of the ground-based microlensing searches, the observations could also be improved, and in this regard a near-IR microlensing campaign from space with the Nancy Grace Roman Space Telescope, formerly known as the Wide-Field InfraRed Survey Telescope or WFIRST is noteworthy \citep{Spergel15,penny19}. The US Decadal Survey recommended the Roman Space Telescope as the top space mission \citep{bla10}.

The present findings suggest a strategy for a Roman microlensing campaign that could complement and augment current ground-based microlensing surveys.
The total area of the Roman footprint is similar to the area covered by VIRCAM at VISTA, which has been able to map efficiently the whole MW bulge and southern disk.
We suggest that a similar mapping pattern be applied across the Galactic plane, covering an area with $|b|<2$ deg. and $|l|<30$ repeatedly.  The cadence should be frequent during the first season (three observations per day per field).
Also, it is of utmost importance that the Roman coverage is extended to the $K_s$ band, as suggested by \cite{Stauffer18}.


If we consider an RC star in the far disk of the Galaxy, at the ``Galactic antipodes" with $R_0=8.3$ kpc and $D=16.6$ kpc from the Sun, a reasonable guess would be a total extinction of $A_V\sim 60$ mag, $A_{K_s} \sim 6$ mag. Such a star would have $m-M=16.0$ mag, and apparent magnitude $K_s\sim 19$ mag. While this is just beyond the detection range for the VVV survey, which has a limiting magnitude $K_s \sim 18.0$  mag typically, it would be readily accessible with the Roman owing to its higher resolution and sensitivity \citep{Spergel15,penny19}. In fact, if the observations are done at $\sim 2 \mu m$, the Roman Telescope offers the possibility of mapping microlensing through the Galactic plane. If the Roman observations are limited to $\leq 1 \mu m$, the rest of the Galaxy will remain hidden and unexplored.

Microlensing observations at low Galactic latitudes are important also to find planets in the bulge. \cite{Henderson14} and \cite{Henderson16} predicted the planet detection fraction as a function of distance considering sources in the bulge, but as an outcome most of the planets would be located in the near disk, with very few of them in the bulge. 
The ability of the near-IR surveys, such as VVV, to see the far-disk sources suggests that the Roman Telescope would efficiently allow discovery of numerous planetary lenses in the bulge.

Note that VIRCAM at the VISTA telescope will be decommissioned after 2020, well before the launch of the Roman Space Telescope, severely hampering the ground-based observational efforts to conduct a microlensing campaign in the near-IR.

In this respect, we have stressed the importance of extending the Roman filter coverage out to $\sim 2 \mu m$ (at the $K_s$ band).
Apparently, current plans do not consider any such near-IR filters  \citep{Green12,Spergel15,penny19}. This limits the Roman Telescope’s vision into the Galactic plane, making it unable to reach microlensing events in the far disk. The gain with longer wavelengths is enormous;  a difference of only one magnitude beyond the bulge provides vast real estate for microlensing in the far disk of our Galaxy.

Also, if we were to see events with sources in the far disk, their relative motions would be larger, and it may be easier in some cases to separate the lens from the source after only a few years. This would allow direct detection of the lens, and better measurement of its physical parameters \citep{Alcock01,gould04}.

\section{Conclusions}
\label{sec:sec6}
We have identified candidate microlensing events with sources in the far disk located in the Galactic plane.
We studied in detail the PMs and CMD of the local areas around the RC sources events using the near-IR photometry from the VVV survey. 
Out of the total sample of $N_{RC}=173$ RC events, we present a list of the best $N=20$  candidate far-disk events, with sources potentially located well beyond the bulge ($D>11$ kpc).
We estimated reddenings and distances for the far-disk RC microlensing event candidates.

This result suggests that as many as $\sim 11 \%$ of RC sources may be located beyond the bulge, in the far disk of our Galaxy.
Our major conclusion is that the events from the far disk could account for a good fraction of the total population of microlensing events in the Galactic plane (in the region $-10.00 \leq l \leq 10.44$, $-0.46  \leq b \leq 0.65 $ deg.).

The mean of the timescale distribution for the far-disk candidates is on average longer ($t_E= 49.3 \pm 7.9$ days) than the value obtained for the Gaussian fit of the bulge RC sources ($t_E = 36.4 \pm 1.1$ days). 
Larger samples such as those likely to be obtained with the Roman Space Telescope in the future would help to confirm this trend.

To accurately measure distances to confirm these candidates we need spectroscopy and higher-resolution observations that can be obtained with ground-based telescopes with adaptive optics and with the next generation of space telescopes. 

We have argued that two major improvements are needed to unveil the structure of the inner MW with microlensing.
First are the models, which need to consider the contribution of events from the far disk.
Second are the observations, which could also be improved, the most promising of which are a near-IR campaign from space with the Nancy Grace Roman Space Telescope. We have stressed the importance of extending the Roman filter coverage out to the $K_s$ band at $\sim 2 \mu m$.
Apparently current plans do not consider this filter, and which is very unfortunate because it would limit the telescope’s vision into the Galactic plane, making it unable to reach microlensing events in the far disk. 

\acknowledgments
We gratefully acknowledge the use of data from the ESO Public Survey program IDs 179.B-2002 taken with the VISTA telescope and data products from the Cambridge Astronomical Survey Unit. 
Support for the authors is provided by the BASAL Center for Astrophysics and Associated Technologies (CATA) through grant AFB 170002, by the Programa Iniciativa Cientifica Milenio grant IC120009, awarded to the Millennium Institute of Astrophysics (MAS), and by Proyecto FONDECYT No. 1170121.



\clearpage

\begin{longtable*}{ l l l l l l l l l l l}
\tablecaption{VVV Survey Far-disk Source Candidates with Their Respective Positions in Galactic Coordinates, $J$ and $K_s$ Magnitudes and the Parameters Obtained Using the Standard Microlensing Model Along with One Standard Deviation Errors.
Typical Positional and Photometric Errors are Provided in the Footnotes. 
} \label{tab:table1} \\
\tablehead{
\colhead{Tile} & \colhead{ID} & \colhead{$l$} & \colhead{$b$} & \colhead{$J$} & \colhead{$K_s$} & \colhead{Amp} & \colhead{$u_{0}$} & \colhead{$t_{0}$} & \colhead{$t_E$} & \colhead{$f_{bl}$}   \\
 & &  {\tiny (deg.)} &  {\tiny (deg.)} & {\tiny (mag)} & {\tiny (mag)}  & &  & {\tiny (MJD)} & {\tiny (days)}  & \\ }
329 & 60206 & -5.97291 & -0.37904 & 17.51 & 15.15 & 0.71 & 0.10 $\pm >$ 2 & 56995.54 $\pm$ 4.34 & 151.28 $\pm$ 49.61 & 0.73 $\pm$ 0.53  \\
329 & 82871 & -5.63381 & -0.42497 & 17.36 & 14.91 & 0.90 & 0.22 $\pm$ 0.07 & 56189.45 $\pm$ 0.70 & 66.31 $\pm$ 13.21 & 0.36 $\pm$ 0.13  \\
331 & 36016 & -3.38748 & 0.21624 & 19.06 & 15.45 & 0.37 & 0.94 $\pm$ 1.58 & 56121.33 $\pm$ 1.71 & 37.63 $\pm$ 42.34 & 1.00 $\pm$ 3.40  \\
331 & 8016 & -3.24922 & 0.63764 & 19.02 & 15.33 & 0.62 & 0.63 $\pm$ 0.93 & 55784.41 $\pm$ 1.20 & 30.68 $\pm$ 29.07 & 1.00 $\pm$ 2.43  \\
331 & 85713 & -2.90247 & -0.25143 & 17.75 & 14.85 & 1.86 & 0.15 $\pm$ 0.02 & 56172.82 $\pm$ 0.12 & 55.75 $\pm$ 4.17 & 0.87 $\pm$ 0.10  \\
331 & 13231 & -3.20217 & -0.10832 & 19.08 & 15.47 & 1.09 & 0.20 $\pm$ 0.20 & 56067.94 $\pm$ 9.06 & 38.11 $\pm$ 16.64 & 1.00 $\pm$ 1.26  \\
332 & 32398 & -2.21056 & 0.22402 & 19.74 & 15.43 & 1.65 & 0.22 $\pm$ 0.12 & 56181.73 $\pm$ 0.43 & 20.99 $\pm$ 7.52 & 1.00 $\pm$ 0.68  \\
332 & 5694 & -1.96859 & -0.27779 & 18.99 & 15.29 & 0.58 & 0.50 $\pm$ 2.61 & 55783.43 $\pm$ 1.18 & 11.21 $\pm$ 25.77 & 1.00 $\pm$ 6.59  \\
332 & 61443 & -1.94419 & 0.16977 & 18.81 & 15.12 & 0.81 & 0.10 $\pm >$ 2 & 56046.33 $\pm$ 1.02 & 64.23 $\pm$ 10.07 & 1.00 $\pm$ 0.29  \\
332 & 90379 & -2.09455 & 0.48129 & 18.16 & 14.94 & 0.82 & 0.17 $\pm$ 0.12 & 55845.35 $\pm$ 1.24 & 68.94 $\pm$ 36.21 & 0.23 $\pm$ 0.19  \\
332 & 90196 & -1.66800 & -0.07156 & 19.82 & 15.24 & 1.02 & 0.33 $\pm$ 0.20 & 56062.40 $\pm$ 2.72 & 74.09 $\pm$ 24.23 & 1.00 $\pm$ 0.73  \\
333 & 25228 & 0.05906 & 0.42445 & 19.02 & 15.24 & 1.76 & 0.10 $\pm >$ 2 & 56525.66 $\pm$ 0.93 & 123.01 $\pm$ 63.35 & 0.26 $\pm$ 0.19  \\
334 & 64986 & 1.59297 & -0.38994 & 17.73 & 14.68 & 0.95 & 0.31 $\pm$ 0.29 & 56516.12 $\pm$ 0.78 & 15.15 $\pm$ 6.48 & 0.87 $\pm$ 0.92  \\
334 & 38331 & 1.44176 & 0.20855 & 19.51 & 15.06 & 0.78 & 0.19 $\pm$ 0.32 & 57179.09 $\pm$ 55.14 & 15.08 $\pm$ 66.67 & 0.54 $\pm$ 8.74  \\
334 & 90309 & 1.41782 & -0.26522 & 19.72 & 15.10 & 1.02 & 0.21 $\pm$ 0.48 & 56190.74 $\pm$ 0.52 & 12.12 $\pm$ 11.26 & 1.00 $\pm$ 1.81  \\
335 & 57964 & 1.99196 & -0.01792 & 19.57 & 15.17 & 0.57 & 0.27 $\pm$ 0.23 & 56177.54 $\pm$ 1.31 & 50.94 $\pm$ 28.54 & 0.26 $\pm$ 0.27  \\
335 & 73192 & 1.84083 & 0.31767 & 19.38 & 15.56 & 1.79 & 0.12 $\pm$ 0.04 & 56851.31 $\pm$ 0.39 & 41.55 $\pm$ 9.71 & 0.58 $\pm$ 0.21  \\
335 & 83301-1647 & 2.98817 & 0.46673 & 18.42 & 15.18 & 1.46 & 0.31 $\pm$ 0.11 & 56840.20 $\pm$ 0.34 & 33.81 $\pm$ 9.08 & 1.00 $\pm$ 0.48  \\
336 & 44838 & 4.19156 & 0.57005 & 17.59 & 14.98 & 2.64 & 0.10 $\pm >$ 2 & 55824.89 $\pm$ 0.28 & 58.33 $\pm$ 9.82 & 1.00 $\pm$ 0.24  \\
337 & 70134 & 5.77744 & -0.24651 & 18.88 & 15.36 & 1.34 & 0.20 $\pm$ 0.08 & 56153.15 $\pm$ 0.27 & 18.21 $\pm$ 5.36 & 0.62 $\pm$ 0.30  \\
\end{longtable*}
\footnotemark{Typical positional errors are 0".1 \citep{Smith17}. } \\
\indent \footnotemark{Typical photometric errors are $\sigma_{K_s} = 0.01$ mag, and $\sigma_{J, H} = 0.03$ mag \citep{saito12, contreras17, alonso18}. } \\

\begin{longtable*}{ l l l l l l l l}
\tablecaption{VVV Survey Far-disk Source Candidates with Their Proper Motions, Centroid Values, Extinction, and Distances. 
Typical Color Errors and Uncertainty in Distances are Provided in the Footnotes. 
} \label{tab:table2} \\
\tablehead{
\colhead{Tile} & \colhead{ID} &  \colhead{$\mu_l$}    &   \colhead{$\mu_b$}     &     \colhead{$J-K_s^{cen}$}    & \colhead{$K_s^{cen}$}       &  \colhead{E($J-K_s$)}   & \colhead{Distance}    \\
& &   {\tiny (mas yr$^{-1}$)}   &     {\tiny (mas yr$^{-1}$)}   &    &  {\tiny (mag)}     & &   {\tiny (kpc)}   \\ }
329 & 60206 & -3.15 $\pm$ 1.17 & 2.12 $\pm$ 1.37 & 2.32 $\pm$ 0.17 & 13.46 $\pm$ 0.74 & 1.72 & 16.6  \\
329 & 82871 & -1.85 $\pm$ 1.73 & 3.94 $\pm$ 1.89 & 2.60 $\pm$ 0.26 & 13.31 $\pm$ 0.60 & 2.00 & 14.1  \\
331 & 36016 & -0.19 $\pm$ 1.25 & 4.13 $\pm$ 1.70 & 3.70 $\pm$ 0.29 & 13.17 $\pm$ 0.66 & 3.10 & 14.4  \\
331 & 8016 & -2.78 $\pm$ 1.50 & -2.95 $\pm$ 1.77 & 2.70 $\pm$ 0.25 & 13.01 $\pm$ 0.69 & 2.10 & 16.7  \\
331 & 85713 & -1.39 $\pm$ 1.16 & -1.50 $\pm$ 1.34 & 2.96 $\pm$ 0.22 & 13.22 $\pm$ 0.69 & 2.36 & 12.7  \\
331 & 13231 & -2.84 $\pm$ 1.64 & -0.42 $\pm$ 1.86 & 3.59 $\pm$ 0.24 & 13.20 $\pm$ 0.69 & 2.99 & 15.0  \\
332 & 32398 & -2.35 $\pm$ 1.54 & 3.31 $\pm$ 1.83 & 3.64 $\pm$ 0.43 & 13.06 $\pm$ 0.74 & 3.04 & 14.5  \\
332 & 5694 & -3.95 $\pm$ 2.17 & -3.62 $\pm$ 1.89 & 3.92 $\pm$ 0.29 & 12.96 $\pm$ 0.60 & 3.32 & 12.9  \\
332 & 61443 & -3.77 $\pm$ 1.59 & 0.59 $\pm$ 1.63 & 3.44 $\pm$ 0.30 & 13.12 $\pm$ 0.82 & 2.84 & 13.1  \\
332 & 90379 & -1.73 $\pm$ 1.20 & 0.12 $\pm$ 1.17 & 2.88 $\pm$ 0.30 & 13.02 $\pm$ 0.67 & 2.28 & 13.4  \\
332 & 90196 & -0.46 $\pm$ 3.03 & -3.86 $\pm$ 2.16 & 3.78 $\pm$ 0.41 & 12.90 $\pm$ 0.60 & 3.18 & 12.9  \\
333 & 25228 & -0.72 $\pm$ 2.71 & 1.14 $\pm$ 2.06 & 3.59 $\pm$ 0.36 & 12.87 $\pm$ 0.53 & 2.99 & 13.4  \\
334 & 64986 & -4.39 $\pm$ 0.78 & 0.29 $\pm$ 0.85 & 3.03 $\pm$ 0.28 & 12.90 $\pm$ 0.65 & 2.43 & 11.6  \\
334 & 38331 & -1.07 $\pm$ 1.48 & -3.34 $\pm$ 1.01 & 3.98 $\pm$ 0.45 & 12.78 $\pm$ 0.46 & 3.38 & 11.5  \\
334 & 90309 & -10.51 $\pm$ 3.47 & -1.30 $\pm$ 3.12 & 3.69 $\pm$ 0.50 & 12.78 $\pm$ 0.53 & 3.09 & 12.3  \\
335 & 57964 & -0.60 $\pm$ 1.66 & 2.95 $\pm$ 1.95 & 3.96 $\pm$ 0.40 & 12.74 $\pm$ 0.41 & 3.36 & 12.1  \\
335 & 73192 & -3.07 $\pm$ 2.22 & 3.10 $\pm$ 1.89 & 3.62 $\pm$ 0.33 & 12.98 $\pm$ 0.61 & 3.02 & 15.4  \\
335 & 83301-1647 & -1.21 $\pm$ 1.48 & -3.69 $\pm$ 2.02 & 2.96 $\pm$ 0.82 & 12.95 $\pm$ 0.63 & 2.36 & 14.8  \\
336 & 44838 & -3.66 $\pm$ 1.24 & -0.59 $\pm$ 1.06 & 2.42 $\pm$ 0.27 & 13.15 $\pm$ 1.17 & 1.82 & 15.0  \\
337 & 70134 & -0.31 $\pm$ 1.67 & -0.58 $\pm$ 1.47 & 3.29 $\pm$ 0.42 & 12.95 $\pm$ 0.79 & 2.69 & 15.1  \\
\end{longtable*}
\footnotemark{Typical color errors are $\sigma_{(J-K_s)_0 }= 0.01 $ mag \citep{alves02} }\\
\indent \footnotemark{According to the error propagation the uncertainty in distance is $\sigma_d \approx 2.9$ kpc. } \\

\end{document}